\documentclass[12pt,aps,prd,preprint,tightenlines,superscriptaddress,showpacs,nofootinbib]
{revtex4-1}
\usepackage{epsfig}
\usepackage{amssymb,amsmath}
\usepackage{hyperref}
\usepackage{color}
\usepackage[utf8]{inputenc} 

\newcommand{\bea}{\begin{eqnarray}}
\newcommand{\eea}{\end{eqnarray}}

\begin{document}
%
\vspace*{1.0cm}

\begin{center}
\baselineskip 20pt 
{\Large\bf 
Experimentally distinguishable origin
for electroweak symmetry breaking
}
\vspace{1cm}

{\large
Victor Baules\footnote{vabaules@crimson.ua.edu} and Nobuchika Okada\footnote{okadan@ua.edu}
}
\vspace{.5cm}

{\baselineskip 20pt \it
Department of Physics and Astronomy, University of Alabama, Tuscaloosa, AL 35487, USA
} 

\vspace{.5cm}

\vspace{1.5cm} {\bf Abstract}
\end{center}

We consider a classically conformal $U(1)$ gauge extension of the Standard Model (SM), 
  in which the $U(1)$ gauge symmetry is radiatively broken by the Coleman-Weinberg mechanism. 
This breaking triggers the electroweak (EW) symmetry breaking through a mixed quartic coupling
  between the $U(1)$ Higgs field and the SM Higgs doublet. 
For two Higgs boson mass eigenstates after the symmetry breaking, 
   $h_1$ (SM-like Higgs boson) and $h_2$ (SM singlet-like Higgs boson), 
   we calculate the Higgs boson trilinear coupling ($g_{h_{1} h_{2} h_{2}}$) in the model 
   by setting the Higgs boson mass spectrum to be $M_{h_1} > 2 M_{h_2}$. 
For a common Higgs mass spectrum and mixing angle between two Higgs fields, 
  we find that $g_{h_{1} h_{2} h_{2}}$ in the classically conformal model
  is highly suppressed compared to that calculated for the conventional Higgs potential,
  where the $U(1)$ and EW symmetry breaking originate from the negative squared masses 
  for the Higgs fields at the tree-level. 
Thus, this coupling suppression is a striking nature of the radiative origin of EW symmetry breaking. 
We then consider how to distinguish this origin at the proposed International Linear Collider (ILC) 
  via precise measurements of anomalous SM Higgs boson couplings and 
  the search for anomalous SM Higgs boson decay $h_1 \rightarrow h_2 h_2$ followed by $h_2 \to b \bar{b}$. 
We conclude that once the anomalous couplings are measured at the ILC, 
  the observation of the anomalous Higgs boson decay is promising in the conventional Higgs potential, 
  while this decay process is highly suppressed and undetectable for the classically conformal model. 


\thispagestyle{empty}

\newpage

\addtocounter{page}{-1}


\section{Introduction} 
\label{sec:1}
The existence of problems not addressed or answered by the Standard Model (SM) has motivated many searches for physics beyond the SM. 
To that end, many extensions of the SM have been proposed, involving various additional particles and/or interactions 
  providing possible solutions to some or all of these problems.
Among these problems is the origin of the electroweak (EW) symmetry breaking.
While successful, there is no compelling explanation for how it arises, 
  since the negative mass-squared term for the Higgs field is generally included by hand.

One interesting way to break a gauge symmetry is the so-called radiative symmetry breaking 
   first proposed by Coleman and Weinberg \cite{Coleman:1973jx}, where classically conformal symmetry is imposed
   at the tree-level scalar potential, and radiative corrections to the potential from the gauge interaction trigger symmetry breaking.
The origin of the symmetry breaking is thus due to quantum corrections, so no negative mass-squared term is added by hand.
However, the Coleman-Weinberg mechanism cannot be naively applied to the SM Higgs sector, 
   as the SM Higgs potential is unbounded from below by quantum corrections from the large top quark Yukawa coupling
   (see for example Ref.~\cite{Sher:1988mj}).    
For this scheme to be implemented, extending the SM is necessary. 
A simple way to do so is to extend the SM with a new $U(1)$ gauge symmetry and to impose classical conformality 
  to forbid explicit mass-squared terms \cite{Hempfling:1996ht, Iso:2009ss}.
The new $U(1)$ gauge symmetry is then broken at some scale $v_{\phi}$ by the Coleman-Weinberg mechanism, where $v_{\phi}$ is the vacuum expectation value (vev) for a Higgs field $\Phi$ for the $U(1)$ gauge symmetry.  
A negative mixing quartic coupling between the SM Higgs doublet and $\Phi$ generates the negative mass squared term 
  for the SM Higgs doublet, driving EW symmetry breaking.
Therefore, the symmetry breaking of the new $U(1)$ gauge symmetry is the sole origin of the mass scale in this model.
This simple mechanism can easily be incorporated into more complex SM extensions.

As the extension includes a new scalar, it is interesting to investigate Higgs boson phenomenology.
We specifically consider the case where the extra Higgs boson is light enough to be produced by SM-like Higgs boson decay.
Such a process originates from the mixing quartic coupling that generates the EW symmetry breaking.
The same coupling also generates a mass mixing between the SM Higgs and $U(1)$ Higgs 
fields.
Due to this mixing, the model predicts anomalous SM Higgs boson couplings.

In this paper, we consider a classically conformal extension to the SM by a (hidden) $U(1)$ gauge group with a Higgs scalar field $\Phi$.
The Coleman-Weinberg mechanism induces radiative symmetry breaking of the new $U(1)$ gauge group, 
   generating the negative mass-squared term for the SM Higgs doublet via a mixed quartic coupling between the SM Higgs and $\Phi$.
This in turn drives the EW symmetry breaking for the SM.
We analyze the Higgs potential and extract a Higgs trilinear coupling $g_{h_1 h_2 h_2}$ 
   between two mass eigenstates, the SM-like Higgs boson $h_1$ and the SM singlet-like Higgs boson $h_2$. 
For comparison, we also analyze the conventional Higgs potential, where the $U(1)$ and EW gauge symmetry breaking
  are triggered by negative squared masses for the two Higgs fields introduced by hand and calculate $g_{h_1 h_2 h_2}$. 
Very interestingly, we find a significant suppression in $|g_{h_1 h_2 h_2}|$ for our classically conformal potential 
  compared to a naively expected value from the conventional Higgs potential 
  while anomalous SM Higgs boson couplings are taken to be the same for both cases.   
We then consider how to experimentally distinguish the radiative breaking origin of the EW symmetry breaking 
  from that in the conventional Higgs potential, namely, a negative mass squared introduced by hand. 
We point out that the future $e^+e^-$ colliders will be a powerful tool for this job 
  by combining a precise measurement of the anomalous SM Higgs boson coupling with the search for 
  an anomalous Higgs boson decay to $h_1 \to h_2 h_2$ followed by $h_2 \to b \bar{b}$. 
If the anomalous SM Higgs boson couplings have been observed, 
  the observation of the anomalous Higgs boson decay is promising for the conventional Higgs potential, 
  while in the classically conformal case, 
  the branching ratio of $h_1 \to h_2 h_2$ is highly suppressed and undetectable. 

%

\section{Classically conformal $U(1)$ extended Standard Model}
\label{sec:2}

In this section, we first consider a minimal extension to the SM involving only an additional $U(1)$ gauge group, which we will denote by $U(1)_{H}$.
In order to focus solely on Higgs phenomenology, we assume in this section that all SM fields to be neutral with respect to $U(1)_{H}$ 
   and no additional chiral fermions are involved in the model. 
We now introduce the $U(1)_{H}$ Higgs field $\Phi$ to have charge $Q_{\Phi} = + 2$ while it is singlet under the SM gauge group. 

Imposing classically conformal invariance, explicit mass terms are forbidden, and the scalar potential at the tree-level is given by
\begin{equation}
\label{treelvV}
 V_{tree} = \lambda_{h}(H^{\dagger}H)^{2} + \lambda_{\phi} (\Phi^{\dagger}\Phi)^{2} - \lambda_{mix}(H^{\dagger}H)(\Phi^{\dagger}\Phi),
\end{equation}
where $H$ is the SM Higgs doublet, and $\lambda_{mix}$ is chosen to be positive.
We make the choice to take $\lambda_{mix} \ll 1$, effectively separating the SM and ``hidden" $U(1)_{H}$ Higgs sectors from each other.
The vacuum of the scalar potential at the tree-level appears at the origin, and EW and $U(1)_{H}$ symmetries are unbroken.
For the $U(1)_{H}$ Higgs sector, we account for the radiative corrections at one-loop level \cite{Coleman:1973jx},
\begin{equation}
V_{1-loop} = 
\frac{\beta_{\phi}}{8}  \left(\textnormal{ln}\left[\frac{\phi^2}{v_{\phi}^{2}}\right]-\frac{25}{6}\right) \phi^{4},
\end{equation}
%
%
where $\phi = \sqrt{2} {\rm Re}[\Phi]$, $\beta_{\phi}$ is given by
\begin{equation}
\beta_{\phi} = \frac{1}{16 \pi^{2}}(20 \lambda_{\phi}^{2} + 96 g_{H}^{4}) \simeq \frac{1}{16 \pi^{2}}( 96 g_{H}^{4}),
\end{equation}
and $g_{H}$ is the gauge coupling of the hidden $U(1)_{H}$.
When these corrections are added to the three-level in Eq.~(\ref{treelvV}), the $U(1)_{H}$ symmetry is radiatively broken 
at $\langle \phi \rangle = v_{\phi}$ as shown in Ref.~\cite{Coleman:1973jx}. 

The full potential is
\begin{equation}
V = \frac{\lambda_{h}}{4} h^{4} + \left[ \frac{\lambda_{\phi}}{4} + \frac{\beta_{\phi}}{8}\left(\textnormal{ln}\left[\frac{\phi^2}{v_{\phi}^{2}}\right]-\frac{25}{6}\right) \right] \phi^{4}- \frac{\lambda_{mix}h^{2}\phi^{2}}{4} 
\end{equation}
where $H = \frac{1}{\sqrt{2}}
\left(
\begin{matrix}
h \\
0
\end{matrix}
\right)$  
with a real scalar $h$ in the unitary gauge.
In the presence of nonzero $\langle \phi \rangle$, the scalar $h$ develops a negative mass-squared term from the mixing quartic coupling.
Subsequently, EW symmetry is broken at $\langle h \rangle = v_{h} = 246$ GeV.

Using the stationary conditions, $\frac{ \partial V }{ \partial \phi} \left|_{\phi = v_{\phi}, \, h=v_{h} }\right. = 0$ and  $\frac{ \partial V }{ \partial h} \left|_{\phi = v_{\phi}, \, h=v_{h} }\right. = 0$, to eliminate $\lambda_{\phi}$ and $\lambda_{mix}$, we express the potential in the following form:
\begin{align}
\begin{split}
V(h,\phi) =& 
\frac{1}{4} 
\left( 
	\frac{m_{h}^{2}}{2 v_{h}^{2}}
\right) h^{4} + \frac{1}{4} 
\left( 
	\frac{11}{6}\beta_{\phi} + \frac{m_{h}^{2}v_{h}^{2}}{2v_{\phi}^{4}}
\right)\phi^{4} 
\\
 + & \frac{\beta_{\phi}}{8}
\left(
	 \textnormal{ln}\left[ 
		\frac{\phi^2}{v_{\phi}^{2}}
	\right] - \frac{25}{6} 
\right)\phi^{4} 
- \frac{1}{4}
\left( 
	\frac{m_{h}^{2}}{v_{\phi}^{2}}
 \right)h^{2}\phi^{2}.
\end{split}
\label{eq:VCW}
\end{align}
where $m_{h}^{2} \equiv 2 \lambda_{h} v_{h}^{2}$.
By shifting the fields $h\rightarrow h + v_{h}$ and $\phi \rightarrow \phi + v_{\phi}$, we can obtain the mass-squared matrix for the physical states,
\begin{equation}
 M_{sq} = 
\begin{pmatrix}
m_{h}^{2} & -M^{2} \\
- M^{2} & m_{\phi}^{2}
\end{pmatrix},
\label{mass-matrix}
\end{equation}
where $M^{2} = m_{h}^{2}\left( \frac{v_h}{v_{\phi}} \right)$ and $m_{\phi}^{2} \equiv  m_{h}^{2} \left( \frac{v_{h}}{v_{\phi}} \right)^{2} + v_{\phi}^{2}\beta_{\phi}$.
We diagonalize $M_{sq}$ by
\begin{align}
\label{diag}
\left(
\begin{matrix}
h \\
\phi
\end{matrix}
\right)
=
\left(
\begin{matrix}
\cos(\theta) & \sin(\theta) \\
-\sin(\theta) &  \cos(\theta)   \\
\end{matrix}
\right)
\left(
\begin{matrix}
h_{1} \\
h_{2}
\end{matrix}
\right),
\end{align}
with a mixing angle $\theta$ defined by $\tan(2\theta) = \frac{2 M^{2}}{m_{h}^{2} - m_{\phi}^{2}}$, 
   where $h_{1}$, $h_{2}$ are the mass eigenstates with mass eigenvalues $M_{h_{1}}$ and $M_{h_{2}}$, respectively.
We see that for a small $\theta$ (or equivalently, a small $\lambda_{mix}$), 
   the SM Higgs is dominated by the $h_{1}$ mass eigenstate and $\phi$  by  $h_{2}$.
In terms of observables ($M_{h_{1}}$, $M_{h_{2}}$, $\theta$), the parameters in Eq.~(\ref{mass-matrix})
can be expressed as follows:
%
\begin{eqnarray}
m_{h}^{2} &=& \frac{1}{2} \left(  M_{h_{1}}^{2} + M_{h_{2}}^{2} + \left(M_{h_{1}}^{2} - M_{h_{2}}^{2} \right) \cos(2 \theta) \right), \\
m_{\phi}^{2} & \equiv &  m_{h}^{2} \left( \frac{v_{h}}{v_{\phi}} \right)^{2} + v_{\phi}^{2}\beta_{\phi}
= \frac{1}{2} \left( M_{h_{1}}^{2} +M_{h_{2}}^{2} - \left(M_{h_{1}}^{2} -M_{h_{2}}^{2} \right) \cos(2 \theta) \right), \\
M^{2} &=& m_{h}^{2}\left( \frac{v_h}{v_{\phi}} \right) =
\frac{1}{2}  \left(M_{h_{1}}^{2} - M_{h_{2}}^{2} \right) \sin(2 \theta) .
\end{eqnarray}
Using these relations, once we fix $M_{h_{1}}$, $M_{h_{2}}$, and $\theta$, the free parameters in the scalar potential of Eq.~(\ref{eq:VCW}), namely $m_h$, $v_\phi$ and $\beta_\phi$, are completely fixed. 
From the above equations, we can see $v_h/v_\phi \simeq \theta$ for $\theta \ll 1$ and $M_{h_{1}}^{2} \gg  M_{h_{2}}^{2}$.

We rewrite the potential in terms of the mass eigenstates $h_{1}$ and $h_{2}$, and extract the trilinear coupling $g_{h_{1} h_{2} h_{2}} \equiv \frac{1}{2} \frac{\partial^{3} V}{\partial h_{1} \partial h_{2}^{2}} \left|_{h_{1} = h_{2} = 0}  \right.$:
\begin{align}
\label{gsimple}
g_{h_{1}h_{2}h_{2}} =& - \frac{M_{h_{2}}^{2} \cos(\theta) \left( M_{h_{1}}^{4} - 5 M_{h_{1}}^{2} M_{h_{2}}^{2} - 2 M_{h_{2}}^{4} + \left( M_{h_{1}}^{4} - 3 M_{h_{1}}^{2} M_{h_{2}}^{2} + 2 M_{h_{2}}^{4}\right) \cos(2 \theta) \right) }{v_{h} \left( M_{h_{1}}^{2} + M_{h_{2}}^{2} + \left( M_{h_{1}}^{2} - M_{h_{2}}^{2} \right) \cos(2 \theta) \right)^{2}} \nonumber \\
& \times \sin^2(\theta).
\end{align}
Assuming $M_{h_{1}} > 2 M_{h_{2}}$ and $\theta \ll  1$, Eq.~(\ref{gsimple}) reduces to
\begin{equation}
\label{CWTC}
g_{h_{1}h_{2}h_{2}} \simeq - \frac{M_{h_{2}}^{2}}{2  v_{h}}  \left(1 - 4 \frac{ M_{h_{2}}^{2}}{ M_{h_{1}}^{2}} \right)  \theta^{2}.
\end{equation}
Note that this trilinear coupling is extremely suppressed compared to the naively expected value derived from the conventional Higgs potential 
   without classical conformal symmetry.
To see this, we now look at the conventional Higgs potential.

%
%


Let us consider the conventional Higgs potential for $h$ and $\phi$ in the unitary gauge\footnote{
For $\hat{v}_\phi \gg v_h$ and $M_{h_{1}} > 2 M_{h_{2}}$, we see $\hat{\lambda}_\phi \ll1$. 
If this is a coupling at the tree-level, the quantum corrections from the $U(1)_H$ gauge interaction is sizable 
and they must be considered. 
Here, we regard the conventional Higgs potential as an effective potential, and all couplings in the potential
are effective couplings after taking quantum corrections into account. }:
\begin{align}
V &= \frac{1}{4} \hat{\lambda}_{h} \big( h^{2} - v_{h}^{2} \big)^{2} + \frac{1}{4} \hat{\lambda}_{\phi} \big(\phi^{2} - \hat{v}_{\phi}^{2} \big)^{2} + \frac{1}{4} \hat{\lambda}_{mix} \big( h^{2} - v_{h}^{2} \big) \big(\phi ^{2} - \hat{v}_{\phi}^{2} \big).
\end{align}
The potential minimum appears at $\langle h \rangle = v_{h}$ and $\langle \phi \rangle = \hat{v}_{\phi}$.
Using the stationary conditions, we rewrite the potential in terms of physical states 
  by shifting $h \rightarrow h + v_{h}$ and $\phi \rightarrow \phi + \hat{v}_{\phi}$:
\begin{align}
\label{eq:potnonconf}
V &= \frac{1}{4} \bigg( \frac{m_{h}^{2}}{2 v_{h}^{2}} \bigg) \big( h^{2} +2 h v_{h} \big)^{2} + \frac{1}{4} \bigg( \frac{m_{\phi}^{2}}{2 \hat{v}_{\phi}^{2}} \bigg) \big( \phi ^{2} + 2 \phi \hat{v}_{\phi} \big)^{2} + \frac{1}{4} \hat{\lambda}_{mix} \big( h ^{2} + 2 h v_{h} \big) \big( \phi ^{2} +2 \phi \hat{v}_{\phi} \big), 
\end{align}
where $m_h^2 =2 \lambda_h v_h^2$, and $m_\phi^2 =2 \hat{\lambda}_\phi \hat{v}_\phi^2$. 
From the above scalar potential we obtain the mass-squared matrix for the conventional case of the form:
\begin{align}
M_{sq} &= 
\begin{pmatrix}
m_{h}^{2} &&M^{2}  \\
M^{2}        &&  m_{\phi}^{2}
\end{pmatrix},
\end{align}
where $M^{2} = \hat{\lambda}_{mix} v_{h} \hat{v}_{\phi}$.
We diagonalize this matrix with Eq.~(\ref{diag}) with the mixing angle defined by $\tan(2 \theta) = - \frac{2 M^{2}}{m_{h}^{2} - m_{\phi}^{2}}$.
In terms of observables $M_{h_{1}}$, $M_{h_{2}}$, $\theta$, and $\hat{v}_{\phi}$, 
we can express the above mass matrix elements as 
\begin{align}
m_{h}^{2} &= \frac{1}{2} \left(  M_{h_{1}}^{2} + M_{h_{2}}^{2} + \left(M_{h_{1}}^{2} - M_{h_{2}}^{2} \right) \cos(2 \theta) \right), \\
m_{\phi}^{2} &= 2 \hat{\lambda}_\phi \hat{v}_\phi^2
= \frac{1}{2} \left( M_{h_{1}}^{2} +M_{h_{2}}^{2} - \left(M_{h_{1}}^{2} -M_{h_{2}}^{2} \right) \cos(2 \theta) \right), \\
M^{2} &= \hat{\lambda}_{mix} v_{h} \hat{v}_{\phi} = -\frac{1}{2}\left(  M_{h_{1}}^{2} -  M_{h_{2}}^{2}\right) \sin(2 \theta).
\end{align}
Note that the conventional Higgs potential of Eq.~(\ref{eq:potnonconf}) is controlled by four free parameters.
In contrast to our conformal model, both $\theta$ and $\hat{v}_{\phi}$ remain free after $M_{h_{1}}$ and $M_{h_{2}}$ are fixed.

Expressing the potential in terms of the mass eigenstates $h_{1,2}$, we extract the trilinear coupling $g_{h_{1} h_{2} h_{2}}$:
\begin{align}
g_{h_{1} h_{2} h_{2}} = \frac{ \left( M_{h_{1}}^{2} + 2 M_{h_{2}}^{2} \right) \left( - v_{h} \cos(\theta) + \hat{v}_{\phi} \sin(\theta) \right) \sin(2 \theta)}{ 2 v_{h} \hat{v}_{\phi} }
\end{align}
This coupling has different behavior for $\theta \ll \frac{v_{h}}{\hat{v}_{\phi}}  \ll 1$ and $\frac{v_{h}}{\hat{v}_{\phi}}  \lesssim \theta  \ll 1$.
For $\theta \ll \frac{v_{h}}{\hat{v}_{\phi}}$, the trilinear coupling $g_{h_{1} h_{2} h_{2}}$ reduces to
\begin{equation}
\label{nonCWTCsmall}
g_{h_{1} h_{2} h_{2}} \simeq - \frac{M_{h_{1}}^{2}}{2 \hat{v}_{\phi}} \left(1 + 2 \frac{M_{h_{2}}^{2}}{M_{h_{1}}^{2}} \right) \theta, 
\end{equation}
while for $ \frac{v_{h}}{\hat{v}_{\phi}} \lesssim \theta $, the coupling is roughly given by
\begin{equation}
\label{nonCWTCbig}
g_{h_{1} h_{2} h_{2}} \simeq  \frac{M_{h_{1}}^{2}}{2 v_{h}} \left(1 + 2 \frac{M_{h_{2}}^{2}}{M_{h_{1}}^{2}} \right) \theta^{2}. 
\end{equation}
Comparing Eqs.~(\ref{nonCWTCsmall}) and (\ref{nonCWTCbig}) with Eq.~(\ref{CWTC}), we see a remarkable difference.
For $\theta \ll \frac{v_{h}}{\hat{v}_{\phi}}$, the coupling from the conformal potential is proportional to $\theta^{2}$, 
while the conventional coupling is proportional to $\theta$ for $\theta \ll \frac{v_{h}}{\hat{v}_{\phi}}$, 
as we would naively expect: $g_{h_{1} h_{2} h_{2}} \sim \hat{\lambda}_{mix} v_{h}$.
This is because in the very small $\theta$ expansion for the trilinear coupling, the lowest order in $\theta$ is cancelled out in the conformal potential.
On the other hand, for $ \frac{v_{h}}{\hat{v}_{\phi}} \lesssim \theta $, the couplings from both potentials have the same $\theta$ dependence, but that of the conformal model still exhibits relative suppression.
The ratio of the trilinear couplings in this region for the same $\theta$ value is found to be 
\begin{equation}
R \equiv \frac{|g_{h_{1} h_{2} h_{2}|_{\rm conf}}|}{|g_{h_{1} h_{2} h_{2}|_{\rm conv}}|}  = \left(\frac{ M_{h_{2}}^{2}}{ M_{h_{1}}^{2} } \right) \frac{  1 - 4 \frac{ M_{h_{2}}^{2}}{ M_{h_{1}}^{2}}}{1 + 2 \frac{M_{h_{2}}^{2}}{M_{h_{1}}^{2}}  } < 0.0505, 
\label{ratio}
\end{equation}
for $M_{h_{2}} < \frac{1}{2} M_{h_{1}}$, with the maximum value obtained when $M_{h_{2}} \simeq 0.33 M_{h_{1}}$. 
One may be interested in setting $\hat{v}_\phi =v_\phi$ in the conventional case, 
although $\hat{v}_\phi$ is a free parameter independent of $\theta$. 
As discussed above, in the conformal model, $\theta \simeq v_h/v_\phi$ for a small $\theta \ll1$. 
Thus, the ratio of the trilinear couplings is given by Eq.~(\ref{ratio}), and the trilinear coupling in the conformal model 
is mush suppressed compared to the one in the conventional case.

%
%
%
%



Note that in both cases, the Higgs anomalous coupling is controlled by $\cos(\theta)$.
This result has interesting implications for Higgs phenomenology, namely, even if the anomalous SM Higgs coupling is detectable in size,
   the SM-like Higgs boson $h_1$ decay to $h_2 h_2$ can be much harder to detect in our model.
This suppression is a characteristic feature of our extended conformally symmetric model, in which EW symmetry breaking is triggered by the radiative $U(1)_{H}$ symmetry breaking.

\section{Numerical Analysis of the Higgs boson trilinear coupling}
\label{sec:3}

\begin{figure}[h!]
\includegraphics[scale=1.0]{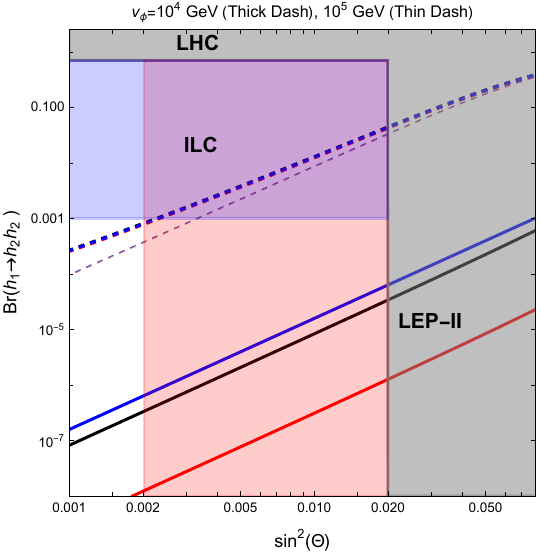}
\caption{Comparison of conventional (dashed lines) and conformal (solid lines) $h_{1} \rightarrow h_{2} h_{2}$ branching ratio at $M_{h_{2}} = 10$ (red), 25 (black), and 50 (blue) GeV. 
Shaded regions correspond to bounds from the proposed ILC, blue for bounds on the branching ratio \cite{Liu:2016zki, Yamamoto:2021kig, Fujii:2015jha}, and red for bounds on $\sin^{2}(\theta)$ \cite{Barklow:2017suo, Bambade:2019fyw}.
The grey shaded regions are excluded by ATLAS($h_{1} \rightarrow h_{2} h_{2}\rightarrow b \bar{b} b \bar{b}$) \cite{ATLAS:2020ahi, Cepeda:2021rql} and LEP-II for $M_{h_{2}} = 25 \textnormal{GeV}$  \cite{Sirunyan:2018koj}.}
\label{fig:BrOfSinsq}
\end{figure}

For an example numerical value, we use the known SM Higgs vev and fix the Higgs mass eigenvalues $M_{h_{1}}$, $M_{h_{2}}$ as follows, alongside a mixing angle $|\theta| = 0.1$,
\begin{align}
M_{h_{1}} = 125 \, \textnormal{GeV}, \quad  M_{h_{2}}= 25 \, \textnormal{GeV}. 
\end{align}
For the classically conformal model, $v_{\phi}$ is totally fixed by the above choices of model parameters to be $v_{\phi} = 2555$ GeV.
We find the triple coupling to be
\begin{equation}
g_{h_{1} h_{2} h_{2}} =  -0.0107,
\end{equation}
while for the conventional Higgs potential, $v_{\phi}$ is a free parameter chosen to be, for example, $v_{\phi} = 10^{4}$ GeV\footnote{
The trilinear coupling weakly depends on $v_\phi$ as shown in Fig.~\ref{fig:BrOfSinsq}.}, and the coupling becomes
\begin{equation}
g_{h_{1} h_{2} h_{2}} = 0.424.
\end{equation}
As explained in the previous section (see Eqs.~(\ref{CWTC}), (\ref{nonCWTCsmall}), and  (\ref{nonCWTCbig})), for $|\theta| = 0.1 \gg \frac{v_{h}}{v_{\phi}}$, the trilinear coupling for the conformal case is suppressed by a factor of $ R \simeq 0.04$.
We use $v_{\phi} = 10^{4}$ GeV as a benchmark value for all calculations for the conventional Higgs potential.

\begin{figure}[h!]
\begin{center}
%
%
\includegraphics[width=0.47\textwidth]{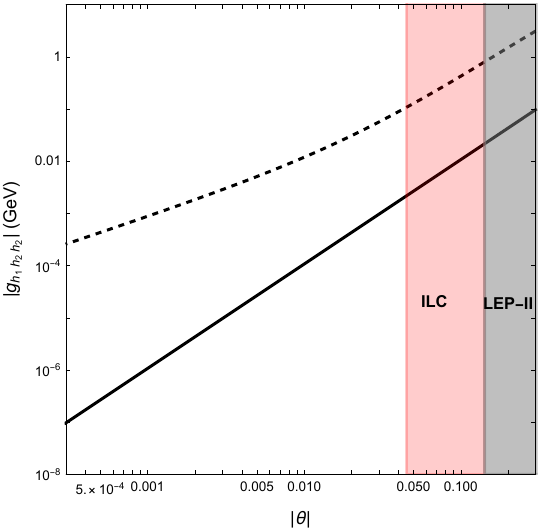} \; 
\includegraphics[width= 0.47\textwidth]{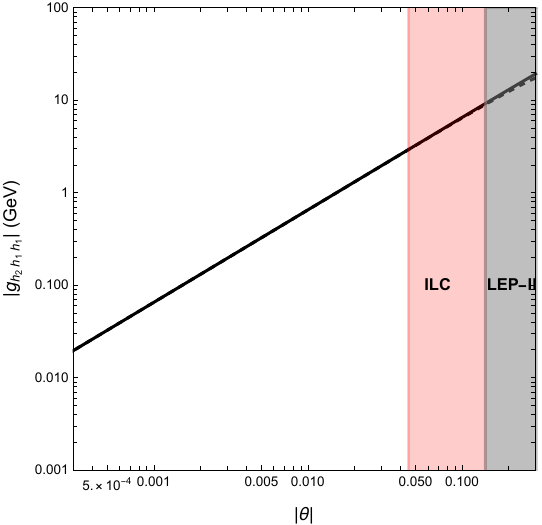}
\caption{ {\it Left Panel}: $|g_{h_{1} h_{2} h_{2}}|$ as a function of $|\theta|$ in our model (solid line) and the conventional Higgs potential (dotted line).
{\it Right Panel}: $|g_{h_{1} h_{1} h_{2}}|$ as a function of $|\theta|$ in our model (solid line) and the conventional Higgs potential (dotted line). 
We set $M_{h_{1}} = 125$ GeV and $M_{h_{2}} = 25$ GeV in our analysis.
 The grey shaded region is excluded by the LEP-II results for the Higgs boson search \cite{Barate:2003sz}, and the red shaded region represents the prospective search reach of the proposed International Linear Collider (ILC)  \cite{Liu:2016zki, Yamamoto:2021kig, Fujii:2015jha}} 
\label{fig:CouplingSupp}
\end{center}
\end{figure}

In the left panel of Fig.~\ref{fig:CouplingSupp}, we plot $|g_{h_{1} h_{2} h_{2}}|$ for the two cases as a function of $|\theta|$.
For $|\theta|\gtrsim \frac{v_{h}}{\hat{v}_{\phi}} \simeq 0.025$, we can see that $|g_{h_{1} h_{2} h_{2}}|$ in the conformal case 
  is suppressed by a factor $R \simeq 0.04$.
As discussed in the previous section (see Eqs.~(\ref{nonCWTCsmall}) and (\ref{nonCWTCbig})),  we can see the transition in the $\sin(\theta)$ dependence from $\sin(\theta)$ to $\sin^{2}(\theta)$ at $|\theta| \simeq \frac{v_{h}}{\hat{v}_{\phi}} \simeq 0.025$.
For $|\theta| \ll \frac{v_{h}}{\hat{v}_{\phi}}$, we can see stronger relative suppression between the couplings, with a suppression factor,
\begin{equation}
\left(  \frac{M_{h_{2}}^{2}}{   M_{h_{1}}^{2}}\right)
\frac{  1 - 4 \frac{ M_{h_{2}}^{2}}{ M_{h_{1}}^{2}}} 
{ 1 + 2 \frac{M_{h_{2}}^{2}}{M_{h_{1}}^{2}}  } 
\left( \frac{ \hat{v}_{\phi}}{ v_{h}}  | \theta| \right)
\ll R.
\end{equation}
The grey shaded region is excluded by the LEP-II results for the Higgs boson search \cite{Barate:2003sz}: Assuming the Higgs boson mass were 25 GeV, the Higgs boson coupling with the $Z$ boson must be suppressed by a factor of $\sin^{2}(\theta) \leq 0.02$.
The red shaded region represents the prospective search reach of the proposed International Linear Collider (ILC), as low as $\sin^{2}(\theta) \simeq 0.002$  \cite{Liu:2016zki, Yamamoto:2021kig, Fujii:2015jha}.
We also calculate the trilinear coupling $g_{h_{1} h_{1} h_{2}}$ with the same parameter choice.
The result is shown in the right panel of Fig.~\ref{fig:CouplingSupp}.
Now, the resultant coupling from our conformal model overlaps well with the same coupling from the conventional Higgs potential, and there is no relative suppression between the two.



We now consider outcomes for experimental probes of our model.
The Higgs physics of interest here are the decay mode $h_{1} \rightarrow h_{2} h_{2}$ and the Higgs anomalous coupling, determined by $\cos(\theta)$.
The partial decay width for the process $h_{1} \rightarrow h_{2} h_{2}$ is given by 
\begin{equation}
\Gamma_{h_{1} \rightarrow h_{2} h_{2}} = \frac{|g_{h_{1} h_{2} h_{2}}|^{2}}{8 \pi M_{h_{1}}} \sqrt{ 1 - 4 \frac{M_{h_{2}}^{2}}{M_{h_{1}}^{2}} }.
\end{equation}
Using the SM-like Higgs boson total decay width to SM particles $\Gamma_{h \rightarrow SM} = 4.07 $ MeV \cite{Dittmaier:2012vm}, the branching ratio to a pair of $h_{2}$ is defined as
\begin{equation}
{\rm Br} \left(h_{1} \rightarrow h_{2} h_{2} \right) = \frac{\Gamma_{h_{1} \rightarrow h_{2} h_{2}}}{\Gamma_{h \rightarrow SM} \cos^2\theta +\Gamma_{h_{1} \rightarrow h_{2} h_{2}}}.
\end{equation}

In Fig.~\ref{fig:BrOfSinsq} we show the results of ${\rm Br} \left(h_{1} \rightarrow h_{2} h_{2} \right)$ 
  for the values of $M_{h_{2}}=$ 10 (red), 25 (black), and 50 (blue) GeV, 
  with solid lines corresponding to the conformal model and dashed lines to the conventional Higgs potential 
  (for which all three lines are nearly overlapping).
As expected from the difference in the couplings, the branching ratios in our conformal model are also extremely suppressed.
The grey shaded region, as before, is excluded by LEP-II and the current bound from the search for 
  $h_{1} \rightarrow h_{2} h_{2} \rightarrow b \bar{b} b \bar{b}$ by the ATLAS collaboration \cite{ATLAS:2020ahi, Cepeda:2021rql}.
Note that since $h_{2}$ is mostly an SM singlet scalar, it will mainly decay to $b \bar{b}$ through the mixing with the SM Higgs doublet.
The prospective ILC search reaches are represented by the blue shaded region 
   corresponding to the branching ratio\cite{Liu:2016zki, Yamamoto:2021kig, Fujii:2015jha}, 
   and red shaded region for the Higgs anomalous coupling \cite{Barklow:2017suo, Bambade:2019fyw}.

The combination of anomalous Higgs decay and anomalous Higgs coupling results will provide a way 
  to distinguish our conformal scenario from the conventional Higgs potential.
For a benchmark value of $\sin(\theta) = 0.1$, for the conventional Higgs case, the anomalous branching ratio 
  and anomalous coupling can both be within the ILC search region, so they can be measured simultaneously.
On the other hand, for our conformal model the branching ratio is highly suppressed, and even if the anomalous coupling were measured, 
  the anomalous decay mode will still evade detection.
This is a way to determine the origin of EW symmetry breaking, i.e. does it occur via the conventional Higgs potential 
  with the negative mass-squared introduced by hand or does it originate from radiative symmetry breaking.

\begin{figure}[t]
\centering
\includegraphics[width=0.47\textwidth]{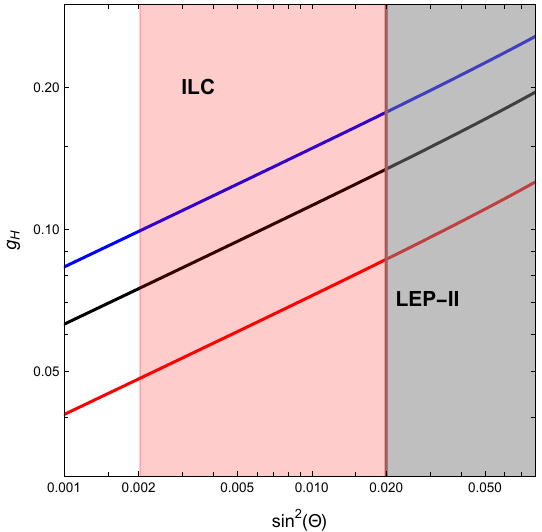} \; 
\includegraphics[width=0.47\textwidth]{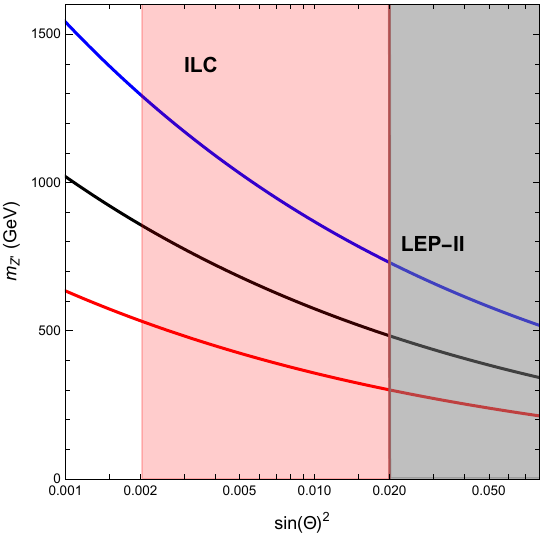}
\caption{{\it Left Panel}: $U(1)_{H}$ gauge coupling ($g_{H}$) as a function of $\sin^{2}(\theta)$.
{\it Right Panel}: $U(1)_{H}$ gauge boson mass ($m_{Z^{\prime}}$) as a function of $\sin^{2}(\theta)$.
The lines correspond to $M_{h_{2}} =$ 10 GeV (red), 25 GeV (black), and 50 GeV (blue).
In both panels, shaded regions correspond to the one in Fig.~\ref{fig:BrOfSinsq}.
}
\label{fig:Zpplots}
\end{figure}

\begin{figure}[t]
\centering
\includegraphics[scale=1.0]{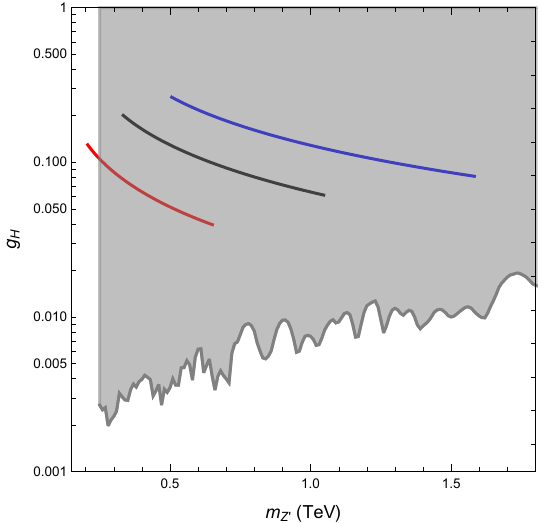}
\caption{The LHC bounds (grey shaded region) on $(m_{Z^{\prime}}$, $g_{H})$ parameter space 
  in the case of $U(1)_{H} = U(1)_{B-L}$, for $M_{h_{2}}=$ 10 GeV (red), 25 GeV (black), and 50 GeV (blue). 
The lines are shown for the range of $0.001 \leq \sin^{2}(\theta) \leq 0.8$, 
 corresponding to the range shown Fig.~\ref{fig:BrOfSinsq}.
}
\label{fig:Zpparamspace}
\end{figure}

In our conformal model, once a value for $M_{h_{2}}$ is fixed, we can determine
   the $U(1)_{H}$ gauge coupling $g_{H}$ and $Z^{\prime}$ boson mass $m_{Z^{\prime}}$ as functions of $|\theta|$.
In Fig.~\ref{fig:Zpplots}, we show the relation between $g_{H}$, $m_{Z^{\prime}}$, and $\sin^{2}(\theta)$
   for $M_{h_{2}} = 10$ GeV, 25 GeV, and 50 GeV, from bottom to top.
If the $Z^\prime$ boson couples to SM fermions, we may consider high-energy collider bounds on the $(m_{Z^\prime}, g_{H})$ parameter space.
As a well known example of such a model, one may consider the minimal $B-L$ (baryon number minus lepton number) model, 
  based on the $U(1)_{B-L}$ gauge group \cite{Davidson:1978pm, Mohapatra:1980qe, Marshak:1979fm,
              Wetterich:1981bx, Masiero:1982fi,  Mohapatra:1982xz, Buchmuller:1991ce}. 
The final results from the LHC Run 2 for the $B-L$ $Z^\prime$ gauge boson resonance search provide us
  with an upper bound on $g_{H}=g_{B-L}$ as a function of $m_{Z^\prime}$, which is presented in Ref.~\cite{Das:2019fee}.
In Fig.~\ref{fig:Zpparamspace}, we show the relationship between $g_{H}$ and $m_{Z'}$, along with the excluded region (grey shaded) 
  by the Large Hadron Collider (LHC) experiments from Ref.~\cite{Das:2019fee}, if we identify our $U(1)_{H}$ with $U(1)_{B-L}$.
The three lines show the relationship for $M_{h_{2}}=$ 10 GeV (red), 25 GeV (black), and 50 GeV (blue) 
  for the range of $0.001 \leq \sin^{2}(\theta) \leq 0.8$, corresponding to the range shown in Fig.~\ref{fig:BrOfSinsq}.
The lines will extend to the right as we take lower values of $\sin^{2}(\theta) < 0.001$. 
To be consistent with the LHC bounds,  $\sin^{2}(\theta)$ must be much smaller than $ 0.001$ (see Fig.~\ref{fig:Zpplots}).
Thus, to explore interesting regions for Higgs phenomenology ($\sin^{2}(\theta) \gtrsim 0.002$), 
  we do not consider the $B-L$ model identification of $U(1)_{H}$.
However, note that the identification of $U(1)_{H}$ with well-known flavor-dependent $U(1)$ extended SMs, 
  such as $U(1)_{B-L_{3}}$ \cite{Babu:2017olk, Alonso:2017uky, Bian:2017rpg} 
  and $U(1)_{L_{\mu} - L_{\tau}}$ \cite{He:1990pn, Foot:1990mn},
  is still possible, as in these models the $Z^{\prime}$ boson has no coupling with the first generation quarks 
  and can easily evade the LHC constraints.

\section{Conclusion and discussion}
\label{sec:4}
We have considered a classically conformal $U(1)$ extension of the SM  
  where the $U(1)$ Higgs field ($\Phi$) with a $U(1)$ charge $+2$ is introduced. 
The $U(1)$ symmetry is radiatively broken by the Coleman-Weinberg mechanism, 
  which generates a negative mass squared to the SM Higgs doublet
  through the mixed quartic coupling in the scalar potential $V \supset - \lambda_{mix} (\Phi^\dagger \Phi) (H^\dagger H)$,
  and as a result, the EW symmetry is broken. 
Therefore, the radiative symmetry breaking of the new U(1) gauge symmetry is the origin
  of the EW symmetry breaking.  
This is a crucial difference from the conventional Higgs potential 
  in which the EW symmetry breaking is triggered by a negative mass squared introduced by hand at the tree-level. 
In order to investigate the Higgs phenomenology of the classically conformal model, we have analyzed the effective 
  Higgs potential and read off the trilinear coupling of the SM-like Higgs boson ($h_1$) 
  with a pair of the SM singlet-like Higgs bosons ($h_2$) for the case of $M_{h_1} > 2 M_{h_2}$. 
For comparison, we have also calculated the trilinear coupling for the conventional Higgs potential. 
What we have found is very intriguing: for the same mixing angle between $H$ and $\Phi$ and 
  the same Higgs boson mass spectrum, the trilinear coupling in the conformal model is highly suppressed 
  compared to the one from the conventional Higgs potential, 
  which is likely a striking nature of the classically conformal potential. 
We then have considered experimental signals for such conformal structure 
   via anomalous SM Higgs boson couplings from the mixing between two Higgs fields  
   and the anomalous (SM-like) Higgs boson decay $h_1 \rightarrow h_2 h_2$ 
   followed by $h_2 \to b \bar{b}$. 
Future $e^+ e^-$ colliders, in particular, the proposed ILC will operate  
  as a Higgs factory which allows us to precisely measure the SM Higgs boson properties.
The result of our analysis indicates that once the anomalous SM-like Higgs boson coupling is measured at the ILC, 
  the observation of anomalous Higgs boson decay $h_1 \to h_2 h_2$ is promising in the conventional Higgs potential, 
  while this decay process is highly suppressed and not detectable in the classically conformal model. 
This is a way to to distinguish the origin of EW symmetry breaking.

As previously discussed, we may extend our classically conformal $U(1)$ model by identifying $U(1)$ 
  with $U(1)_{B-L_{3}}$ or $U(1)_{L_{\mu} - L_{\tau}}$. 
Another interesting possibility of such an extension is to consider the $U(1)$ sector 
  as a ``Dark Sector" which supplements the SM with a dark matter candidate. 
Note that although the $Z^\prime$ gauge boson can generally have a kinetic mixing 
  with the SM hyper-charge gauge boson, if the kinetic mixing is tuned to be zero, 
  the $Z^\prime$ gauge boson is stable and hence a dark matter candidate. 
Therefore, the classically conformal gauged $U(1)$ extension can not only provide the SM
  with the dynamical origin of the EW symmetry breaking but also supplement the SM 
  with the $Z^\prime$ boson dark matter. 
Based on a new $SU(2)$ gauge symmetry, such a model has already been proposed in Ref.~\cite{Hambye:2013dgv},
  although the Higgs boson phenomenology that we have investigated in this paper is not addressed. 
In the present paper, we have considered the $U(1)$ gauge theory, but extending it to a non-Abelian 
  gauge group is straightforward and what we have investigated in this paper 
  remains essentially the same for such extension. 
It is worth investigating the Dark Sector in light of complementarity between dark matter physics
  and Higgs boson phenomenology at future collider experiments \cite{future:2022}.

\section*{Acknowledgments}
The authors V.B.~and N.O.~would like to thank Digesh Raut for useful discussions. 
This work is supported in part by the GEM Science Ph.D. Fellowship (V.B.) and the United States Department of Energy Grants DE-SC-0012447 (N.O.).

\bibliographystyle{utphysII}
\bibliography{References}

\end{document}